\documentclass[12pt]{article}

\usepackage{amsmath,amsfonts,amssymb}
\usepackage{graphicx}
\usepackage{psfrag}
\usepackage{enumerate}
\usepackage{mathrsfs}
\usepackage{cite}

\makeatletter\renewcommand\section{\@startsection {section}{1}{\z@}%
                                   {-3.5ex \@plus -1ex \@minus -.2ex}
                                   {2.3ex \@plus.2ex}%
                                   {\normalfont\large\bfseries}}
\renewcommand\subsection{\@startsection{subsection}{2}{\z@}%
                                     {-3.25ex\@plus -1ex \@minus -.2ex}%
                                     {1.5ex \@plus .2ex}%
                                     {\normalfont\bfseries}}

\newcommand{\be}{\begin{equation}}
\newcommand{\ee}{\end{equation}}

\newcommand{\eeq}{\end{eqnarray}}

\def\[{\left [}
\def\]{\right ]}
\def\({\left (}
\def\){\right )}

\def\r2{\sqrt{2}}

\newcommand{\bbibitem}[1]{\bibitem{#1}\marginpar{#1}}

\def\Label#1{\label{#1}%
  \smash{\hbox to0pt{\raise1ex\hbox{\tiny[#1]}\hss}}}
\def\noLabels{\let\Label=\label}
\def\nobbibitem{\let\bbibitem=\bibitem}

\begin{document}
\noLabels 
\nobbibitem 




\begin{center}

{\Large \bf Barnacles -- A Novel Channel for Vacuum Decay}

\vspace{5mm}

Bart{\l}omiej Czech


\vspace{5mm}

\bigskip\centerline{
\it Department of Physics and Astronomy}
\smallskip\centerline{\it University of British Columbia}
\smallskip\centerline{\it 6224 Agricultural Road}
\smallskip\centerline{\it Vancouver, BC V6T 1Z1, Canada}
\bigskip\medskip


\end{center}
\setcounter{footnote}{0}
\begin{abstract}
\noindent
I show that cosmological bubble walls in the thin wall approximation are unstable to the creation of ``barnacles'' -- loci of different wall tension adjacent to regions filled with a third vacuum. Barnacle formation leads to the same observational consequences as the extensively studied bubble collision scenario, but occurs exponentially more often. The process is described by a saddle point of the thin wall action with two negative modes.
\end{abstract}

\newpage


The eternal inflation paradigm posits that bubbles occupied by different vacua continuously nucleate, expand, and collide with each other amid an eternally inflating false vacuum. In the thin wall approximation \cite{Coleman:1977py, Callan:1977pt, colemanbook}, a nucleation event is described by a saddle point of the Euclidean action
\begin{equation}
S = \int d^4 x \big( -\epsilon + \sigma \delta({\rm walls}) \big)
\label{action0}
\end{equation}
with a single negative mode. Here $\epsilon$ is the energy density and $\sigma$ denotes the tension of a thin wall, whose contribution is confined to the world-volume selected by the $\delta$-function. Generically, more than one generation of bubbles is expected to occur: starting in a false vacuum, typically a (first generation) bubble of intermediate vacuum will host a number of (second generation) bubbles of true vacuum. Such histories are associated with saddle points of (\ref{action0}) with two negative modes, one for each nucleation.

This paper points out that the cylindrically symmetric saddle points discovered in \cite{barnacles} mediate a novel class of second generation nucleation events -- the formation of ``barnacles.'' These objects are loci of different wall tension that are adjacent to regions filled with a third vacuum; they can nucleate on walls of first generation bubbles. In what follows I explain how to compute the rate of barnacle formation. I also point out that barnacle production leads to the same observational consequences as the extensively studied bubble collision scenario \cite{fhs, aguirre1, worldscollide, aguirre2, worldscollide2, aguirre3, disintegration, flow, kinetictr, polarization, polarization2} (see \cite{mattsreview} for a recent review), but is far more generic: the expected number of barnacles in our past lightcone is exponentially larger than the expected number of collisions when these quantities can be meaningfully compared.

Consider a potential landscape containing at least three nondegenerate vacua: a false vacuum A, an intermediate vacuum B and a true vacuum C. Denote the respective vacuum energies $\epsilon_A$ etc. and vacuum energy differences $\epsilon_{AB}$ etc. Working in the thin wall approximation, one defines three wall tensions $\sigma_{AB}, \sigma_{AC}, \sigma_{BC}$ and writes down action (\ref{action0}). When the vacua B and C decouple, that is when $\sigma_{AB}, \sigma_{AC} \to \infty$, the Euclidean partition function for a false vacuum region of volume $V$ is simply:
\begin{equation}
\mathcal{Z} = e^{-S} = e^{\epsilon_A V}
\end{equation}
When transitions to vacuum B are turned on but C remains inaccessible ($\sigma_{AB} < \infty = \sigma_{AC} = \sigma_{BC}$), the partition function includes contributions from saddle points with a single negative mode\footnote{I ignore upward transitions, which are not relevant to the purposes of this paper.}. These are the familiar spherical instantons portrayed in the left column of Table~\ref{saddles}. Summing over all $p$-instanton sectors, one gets:
\begin{equation}
\mathcal{Z} = e^{\epsilon_A V} \sum_p \frac{(iK_{AB}e^{-S_{AB}}V)^p }{p!} = e^{(\epsilon_A + iK_{AB}e^{-S_{AB}})V} \equiv e^{\tilde{\epsilon}_A V}
\label{pathab}
\end{equation}
Here $iK_{AB}$ is a determinant factor coming from a saddle point evaluation of the path integral. It is imaginary, because the saddle point has a single negative mode. The action of the instanton is given by
\begin{equation}
S_{AB} = -\epsilon_{AB} {\rm vol}(B_4) + \sigma_{AB} {\rm vol}(S_3),
\label{single}
\end{equation}
with the volumes of the four-dimensional ball and three-dimensional sphere evaluated at the critical radius
\begin{equation}
R_{AB} = \frac{3\sigma_{AB}}{\epsilon_{AB}}.
\end{equation}
Overall, the net effect of nucleations of B-bubbles is to shift the effective vacuum energy density of A by an imaginary term, interpreted as the rate of decay of vacuum A:
\begin{equation}
\epsilon_A \to \tilde{\epsilon}_A = \epsilon_A + iK_{AB}e^{-S_{AB}} \equiv \epsilon_A + i \Gamma_{AB}
\label{ashift}
\end{equation}

\begin{table}[!t]
\begin{center}
\begin{tabular}{|c|c|c|c|}
\hline
\# & $\sigma_{AB} < \infty$ & $\sigma_{BC} < \infty$ & $\sigma_{AC} < \infty$
\\ \hline

1 &
\parbox{3.4cm}{\begin{center}
\raisebox{-0.4in}[0.3in][0.3in]{\includegraphics[scale=0.2]{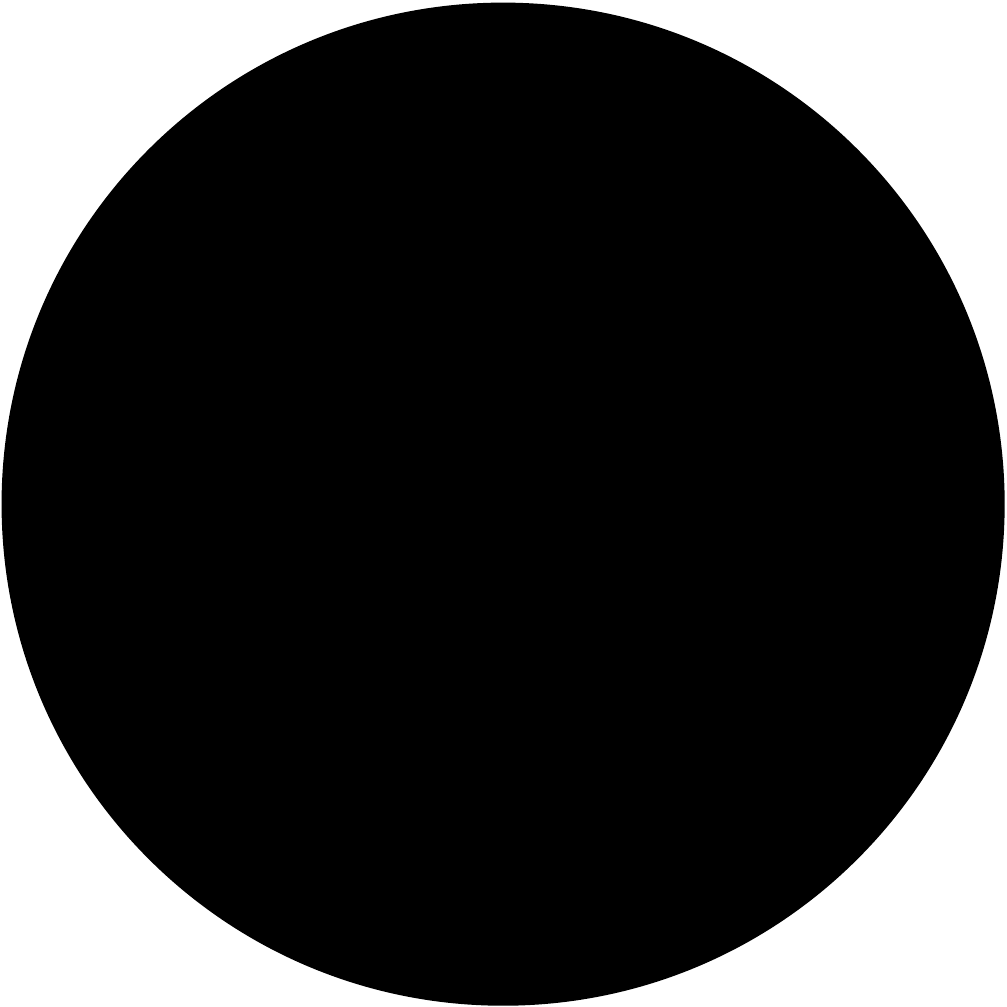}} 
\end{center}}
&
&
\parbox{3.4cm}{\begin{center}
\raisebox{-0.3in}[0.3in][0.3in]{\includegraphics[scale=0.15]{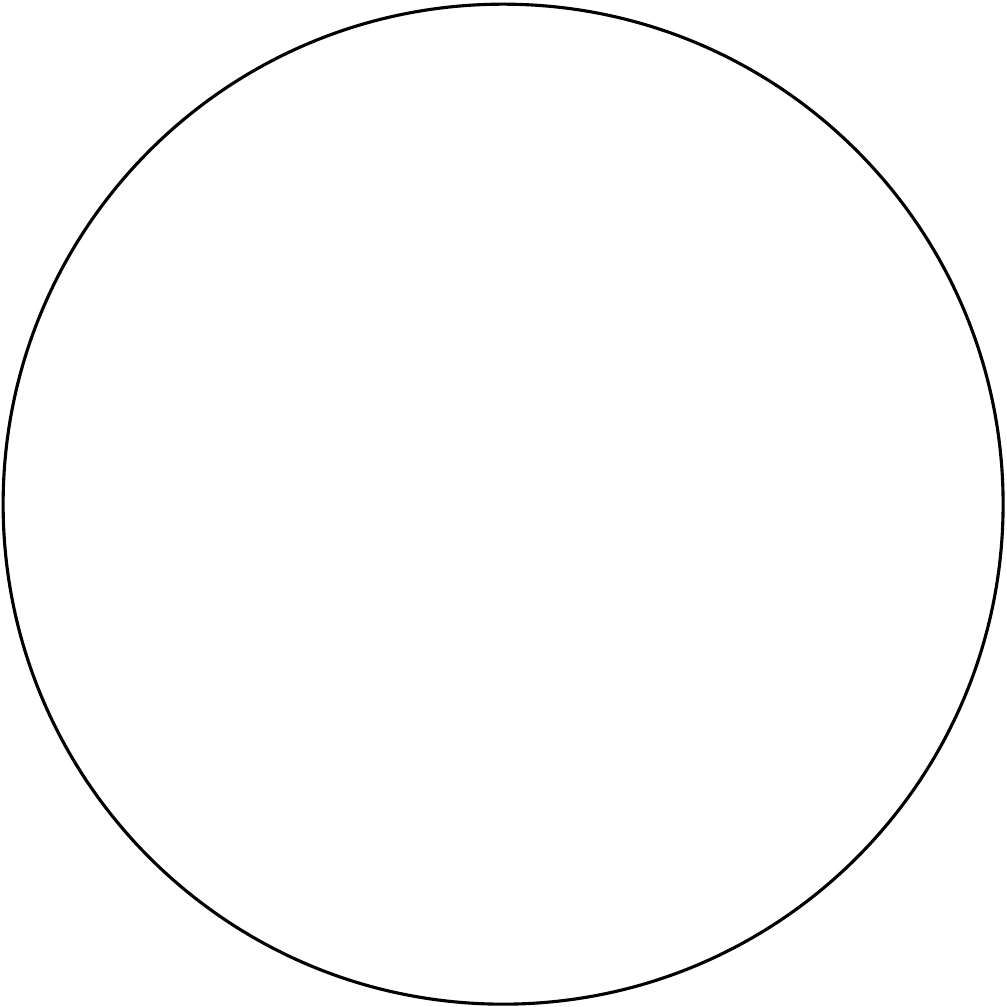}} 
\end{center}}
\\ \hline

2 &
&
\parbox{3.4cm}{\begin{center}
\raisebox{-0.4in}[0.3in][0.3in]{\includegraphics[scale=0.2]{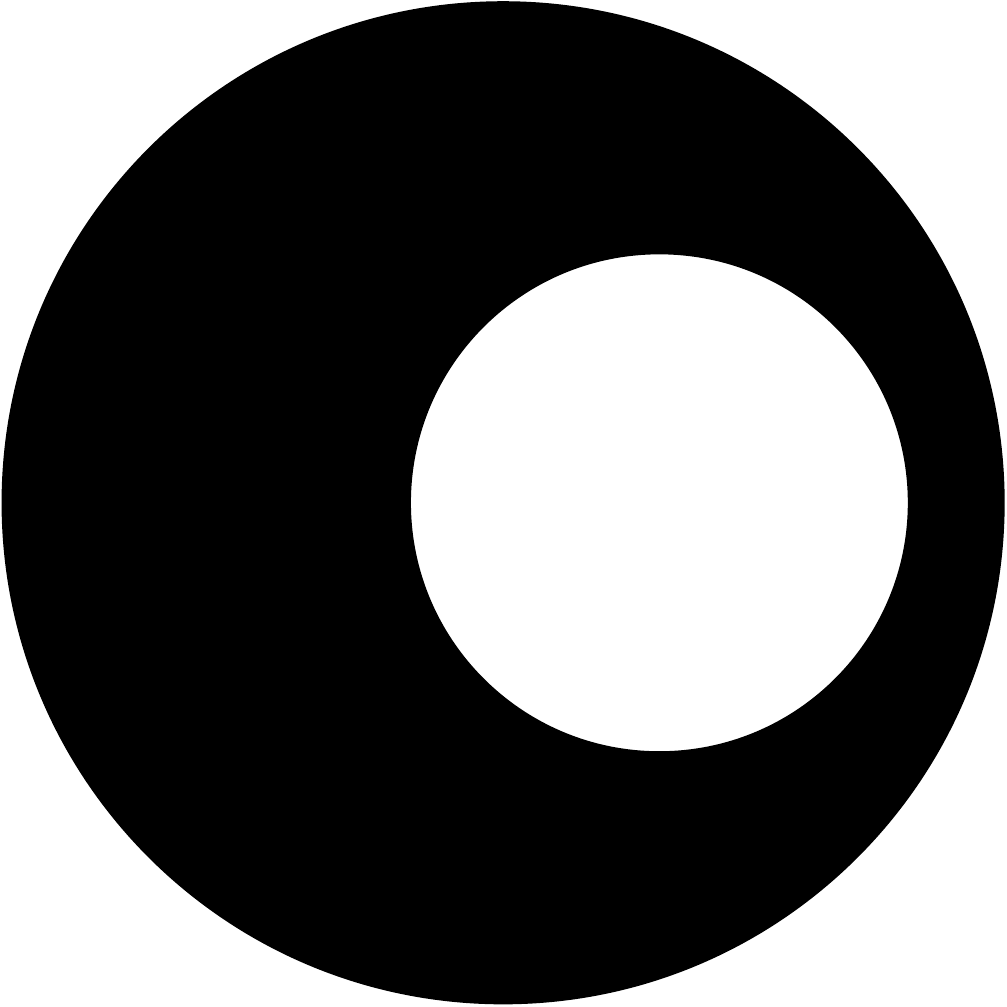}} 
\end{center}}
&
\parbox{3.4cm}{\begin{center}
\raisebox{-0.4in}[0.3in][0.3in]{\includegraphics[scale=0.2]{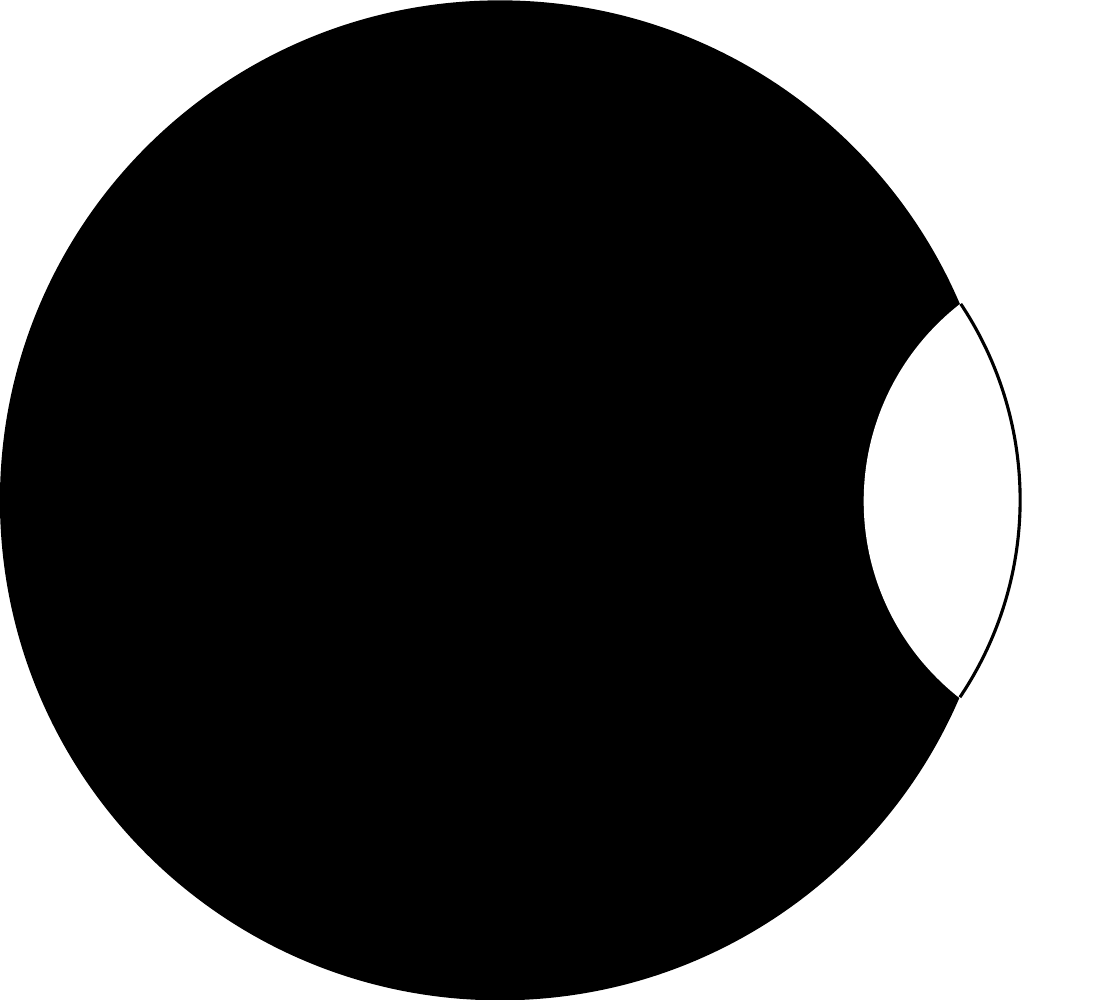}} 
\end{center}}
\\ \hline

\end{tabular}
\end{center}
\caption{Saddle points of the thin wall action, which appear as successive wall tensions are set to finite values, organized by the number of negative modes, i.e. the generation to which the corresponding decay channel belongs. Closed regions drawn in black / white are filled with intermediate (B) / true (C) vacuum; the outside region is the false vacuum (A).}
\Label{saddles}
\end{table}

Next turn on B$\to$C nucleations, setting $\sigma_{BC}$ to a finite value, but still keeping $\sigma_{AC}$ infinite. Now the path integral admits saddle points with two negative modes, portrayed in the center column of Table~\ref{saddles}. When we include their contribution in the path integral, each spherical instanton of vacuum B comes dressed with balls of vacuum C contained inside it. When we account for this, the exponentiated term in eq.~(\ref{pathab}) gets shifted by:
\begin{eqnarray}
iK_{AB}e^{-S_{AB}}V 
& \to & 
iK_{AB}e^{-S_{AB}}V\sum_m\frac{(iK_{BC}e^{-S_{BC}}{\rm vol}(B_4)\vert_{R_{AB}})^m}{m!} 
\nonumber \\
& = & 
iK_{AB} V \exp{\left(-S_{AB} + iK_{BC}e^{-S_{BC}}{\rm vol}(B_4)\vert_{R_{AB}}\right)}
\label{bshift}
\end{eqnarray}
The index $m$ counts C-balls, each of which comes with the factor ${\rm vol}(B_4)\vert_{R_{AB}}$ from integrating over the translational zero modes. Comparing with eq.~(\ref{single}), we see that the net effect of including saddle points with two negative eigenvalues is to induce a shift:
\begin{equation}
\epsilon_B \to \tilde{\epsilon}_B = \epsilon_B +  iK_{BC}e^{-S_{BC}} \equiv \epsilon_B + i \Gamma_{BC}
\label{gammabc}
\end{equation}
Of course, this shift can never appreciably affect the decay rate of A. However, from the viewpoint of an observer living inside a first generation bubble B, the shift is of the same form as (\ref{ashift}) and corresponds to the leading decay channel of her ambient vacuum.

Our goal is an analogous understanding of what happens when the third wall tension, $\sigma_{AC}$, becomes finite. This introduces two types of saddle points, shown in the right column of Table~\ref{saddles}. The first is a familiar, spherical instanton of vacuum C in A, which has a single negative mode and represents the production of first generation C-bubbles. The second type is cylindrically symmetric and has two negative modes; its existence was reported in \cite{barnacles}. 

To interpret this novel object, let us proceed by analogy with the other saddle point with two negative modes, which was interpreted in eqs.~(\ref{bshift}, \ref{gammabc}). A second generation C-bubble could be placed anywhere in the volume of the B-bubble, so it came with a factor of ${\rm vol}(B_4)\vert_{R_{AB}}$. The feature represented by our novel saddle point can be placed anywhere on the surface of the wall of a first generation bubble. We call it a ``barnacle.'' From zero mode integration, every barnacle will come with a factor of ${\rm vol}(S_3)\vert_{R_{AB}}$. Summing over all $n$-barnacle sectors and keeping the $m$ second generation C-bubbles inside B included in eq.~(\ref{bshift}), the exponential term in (\ref{pathab}) shifts by:
\begin{eqnarray}
iK_{AB}e^{-S_{AB}}V 
& \to & 
iK_{AB}e^{-S_{AB}}V
\sum_m\frac{(iK_{BC}e^{-S_{BC}}{\rm vol}(B_4)\vert_{R_{AB}})^m}{m!} 
\sum_n\frac{(iK_{b}e^{-S_{b}}{\rm vol}(S_3)\vert_{R_{AB}})^n}{n!}
\nonumber \\
& = & 
iK_{AB} V \exp{\left( -S_{AB} + iK_{BC}e^{-S_{BC}}{\rm vol}(B_4)\vert_{R_{AB}} + iK_{b}e^{-S_{b}}{\rm vol}(S_3)\vert_{R_{AB}} \right)}
\label{barnacleshift}
\end{eqnarray}
Here $S_b$ is the increase in action when the wall of a B-bubble is dressed with a single barnacle, and $i K_b$ is the corresponding ratio of determinant factors (which is imaginary, because incorporating one barnacle introduces one extra negative mode). Evidently, the nonspherical saddle points also induce a shift in the effective thin wall parameters, but this time the shifted quantity is the wall tension:
\begin{equation}
\sigma_{AB} \to \tilde{\sigma}_{AB} = \sigma_{AB} - iK_{b}e^{-S_{b}} \equiv \sigma_{AB} - i \Gamma_b 
\label{walldecay}
\end{equation}
The last step makes the natural identification of the imaginary part of the tension as the wall$\to$barnacles decay rate per wall three-volume. One might worry that this computation is only valid for barnacles whose creation is simultaneous with the nucleation of the first generation B-bubble, because only then is it possible to analytically continue the action to arrive at the Euclidean path integral (\ref{pathab}) shifted by (\ref{barnacleshift}). But the $SO(3,1)$ symmetry of the B-bubble guarantees that if eq.~(\ref{walldecay}) applies anywhere, then it applies everywhere on the wall.

As with second generation C-bubbles, the appearance of barnacles does not affect the decay rate of vacuum A except formally. However, for an observer inside a first generation B-bubble, barnacle formation represents a very relevant instability of the world around her. It is distinct from the instability of the B-vacuum due to formation of C-bubbles, but it is on the same footing with it. To appreciate this, it is useful to understand the Lorentzian evolution of a barnacle formed on a wall separating vacua A and B. The detailed shape of the barnacle depends on the landscape parameters (Types I, II, III in Table~\ref{allsols}), but its general characteristics do not: the barnacle looks like a small, axisymmetric bubble of vacuum C, which spontaneously pops into existence ``inside'' the AB wall. It is surrounded by thin walls of the BC and AC types, initially at rest. Because the C-vacuum has lower energy density than A and B (by assumption), this bubble will expand into its two neighbors with a speed asymptoting to the speed of light. The Lorentian evolution of the two walls is described by three-dimensional hyperboloids, the analytic continuation of the sphere slices making up the Euclidean saddle point. These two hyperboloids are constrained to coalesce with the hyperboloid of the original AB-wall. The locus where the three walls meet, called the junction in \cite{barnacles}, is itself a two-dimensional hyperboloid, because its Euclidean precursor is a two-dimensional sphere as a consequence of cylindrical symmetry.

\begin{table}[!t]
\begin{center}
\begin{tabular}{|c|c|c|c|c|}
\hline

\parbox{1.4in}{\begin{center} Type I: \\ \rule{0pt}{3ex} lives on AB-walls \end{center}}
& 
\parbox{3.4cm}{\begin{center}
\raisebox{-0.4in}[0.3in][0.2in]{\includegraphics[scale=0.2]{mmp.pdf}} 
\end{center}}
& & 
\parbox{1.4in}{\begin{center} Type III: \\ \rule{0pt}{3ex} lives on AB-walls and AC-walls \end{center}}
& \raisebox{-0.4in}[0.5in][0.5in]{\includegraphics[scale=0.2]{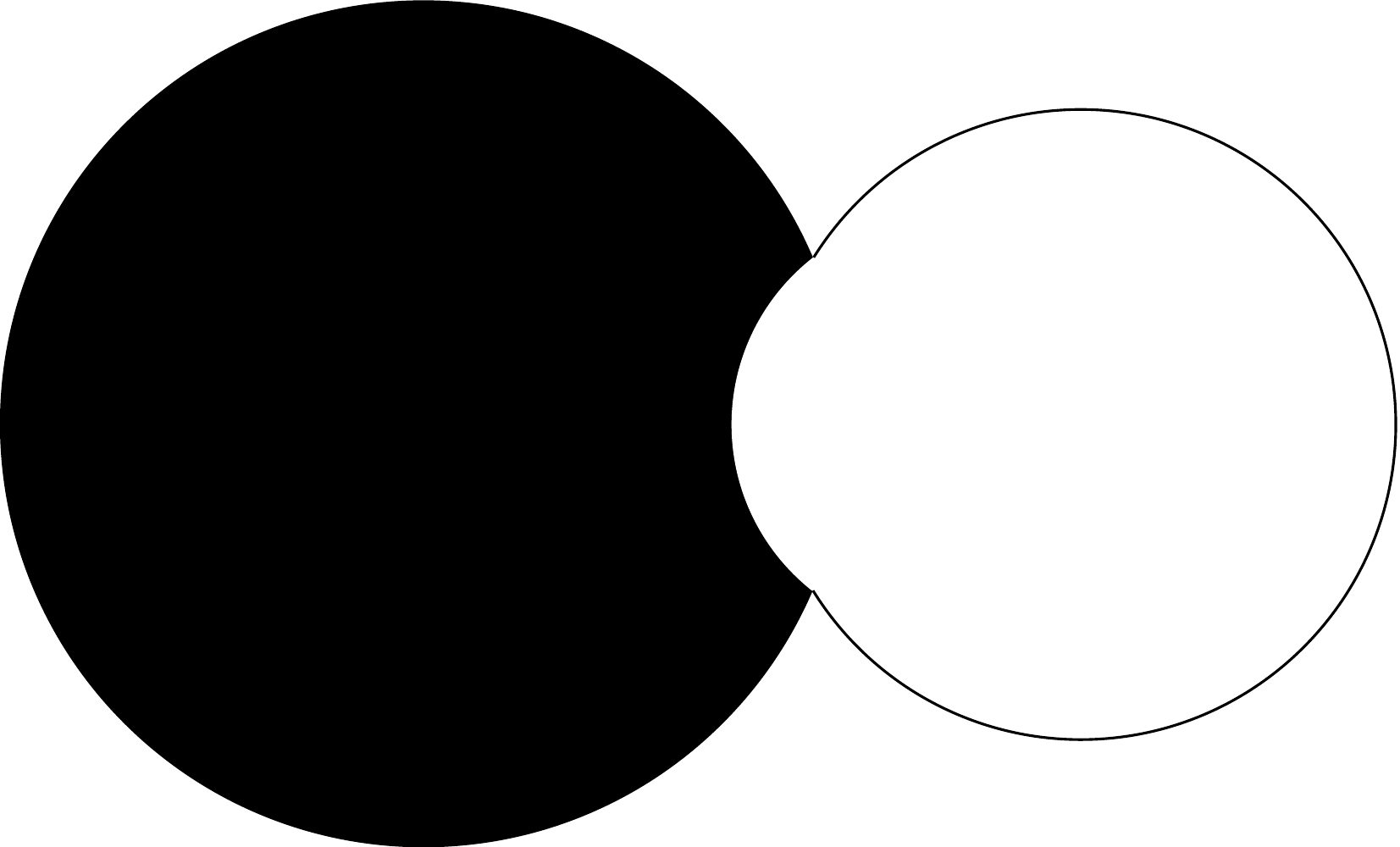}} 
\\ \hline

\parbox{1.4in}{\begin{center} Type II: \\ \rule{0pt}{3ex} lives on AB-walls \end{center}}
& \raisebox{-0.4in}[0.5in][0.5in]{\includegraphics[scale=0.2]{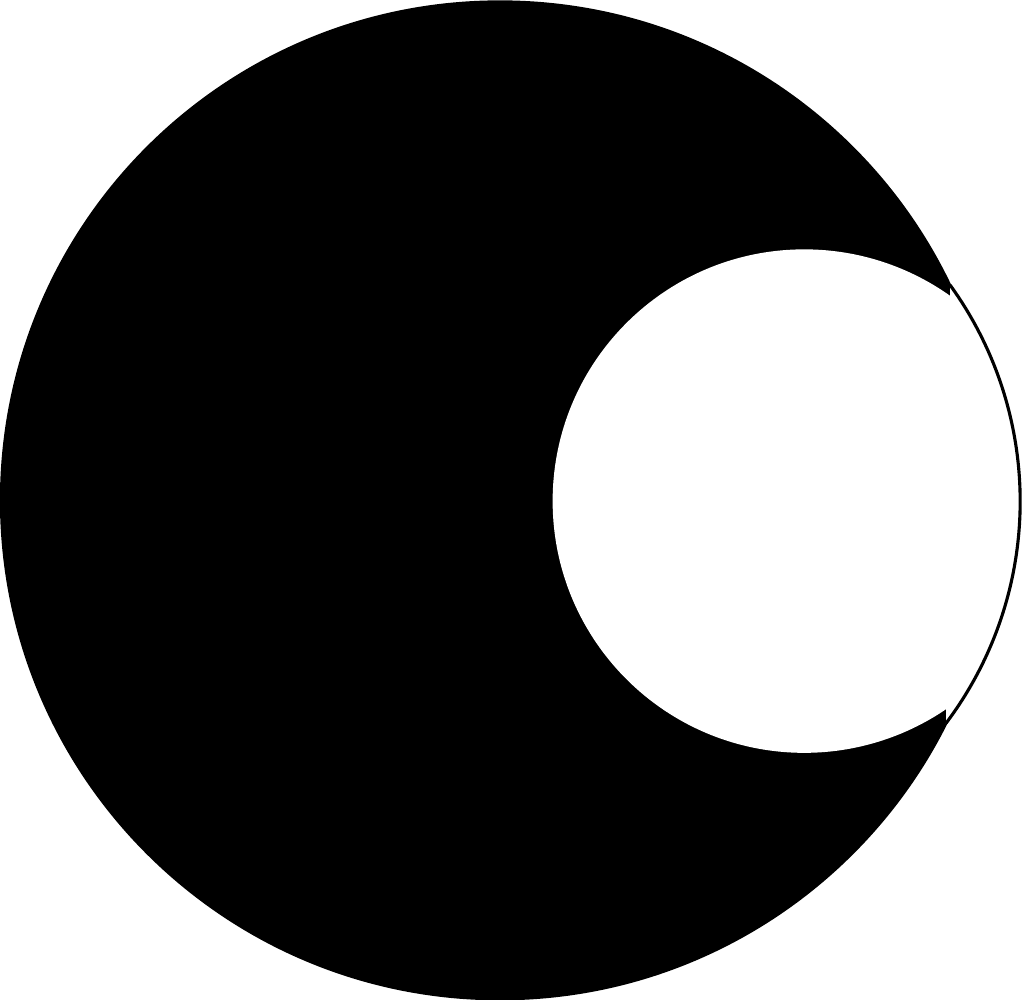}} 
& & 
\parbox{1.4in}{\begin{center} Type IV: \\ \rule{0pt}{3ex} lives on AC-walls \end{center}}
& \raisebox{-0.3in}[0.4in][0.4in]{\includegraphics[scale=0.2]{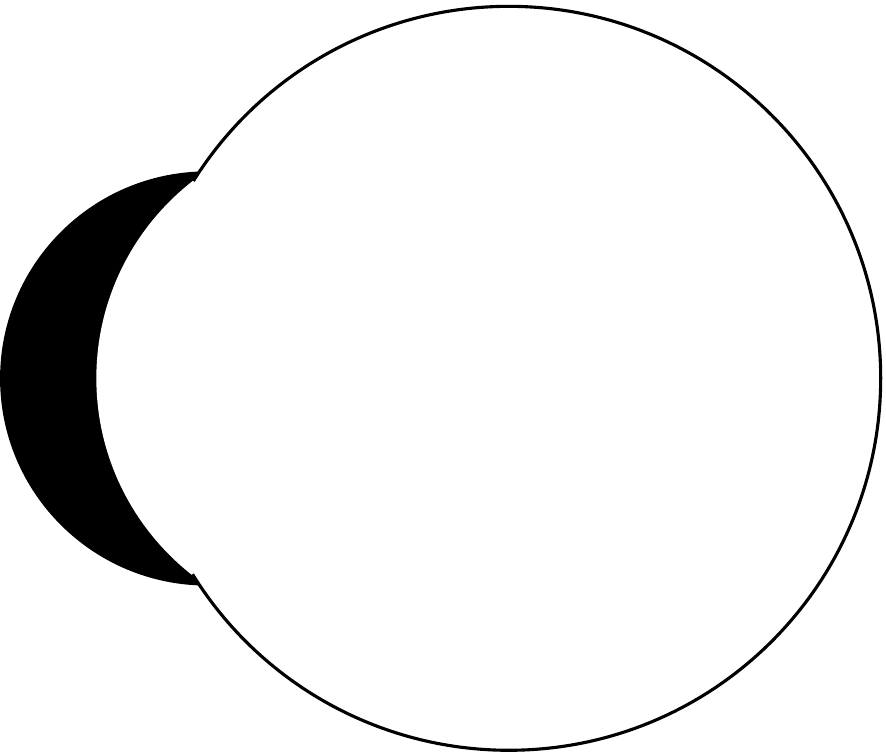}} 
\\ \hline

\end{tabular}
\end{center}
\caption{Taxonomy of barnacles. Closed regions drawn in black / white are filled with intermediate (B) / true (C) vacuum; the outside region is the false vacuum (A). For bubbles of true vacuum, barnacles of Types III and IV cover a small part of the surrounding wall while barnacles of Types I and II cover half or more of the surrounding wall. For a given choice of landscape parameters, only one type of barnacle exists; see \cite{barnacles} for more details.}
\Label{allsols}
\end{table}

Barnacles of Types III and IV (see Table~\ref{allsols}) may also pop up on walls separating the true and the false vacuum (as opposed to true and intermediate, discussed above). In this case, they are filled with the intermediate vacuum and therefore their walls accelerate away from the center of the true vacuum bubble. Indeed, for an observer living in the true vacuum C, the only instabilities of the world around her are (a) barnacles on the walls surrounding her bubble, and (b) nucleations of new bubbles in the mother vacuum, which may subsequently collide with her bubble. The observational consequences of the latter scenario (if we live in a bubble of vacuum C) have been studied in \cite{worldscollide2, flow, polarization, polarization2}. As soon as we realize the importance of barnacles, the first question is: how do the observational signatures of barnacle formation differ from a bubble collision?

The answer is that they do not differ at all. The analyses of \cite{worldscollide2, flow, polarization, polarization2} rely on a single starting point: that the wall surrounding our bubble contains an $SO(2,1)$-invariant patch of different wall tension, which accelerates away from the center of our bubble. All observational consequences of bubble collisions -- from a CMB cold / hot spot \cite{worldscollide2} and its associated polarization pattern \cite{polarization, polarization2} to galaxy flow \cite{flow}, follow from this single input. But this input is shared by the barnacle: it also leads to an $SO(2,1)$-invariant patch of different wall tension, which accelerates away from us whenever a collision-formed domain wall would. Thus, the hitherto reported observational signatures of bubble collisions apply in their entirety to barnacle formation.

This motivates the next question: which is more generic? The existence of barnacles relies on there being at least one extra vacuum, in addition to the two vacua inside and outside the observer's bubble. In contrast, bubble collisions occur also in the absence of other vacua in the landscape -- in which case the colliding bubbles are filled with the same vacuum and no domain wall forms between them.

I now argue that if a third vacuum exists, a barnacle scenario is far more generic than the bubble collision scenario. I need two preliminaries for this discussion. The first is that consistency with observations requires all domain walls to accelerate away from us, so our vacuum must be the true vacuum C. The second is that for any given choice of landscape parameters, precisely one of the four types of barnacles displayed in Table~\ref{allsols} exists \cite{barnacles}. I begin by focusing on the part of parameter space, where barnacles are of Types III or IV. In this case it is possible to think of barnacles as populating the surface of the wall of our spherical, C-vacuum bubble. It is then meaningful to ask how many barnacles are expected to be visible in our past lightcone and how that number compares with the expected count of visible collision events.

Ref.~\cite{disintegration} computed the expected number of bubble collisions in our past lightcone. In terms of quantities defined in the present paper, their result reads
\begin{equation}
N_c \sim (\Gamma_{AB} + \Gamma_{AC}) \frac{A}{G_N \epsilon_A},
\label{collisioncount}
\end{equation}
where $G_N$ is Newton's constant. The $A$ denotes the two-dimensional cross-sectional area of wall intersecting our past lightcone (Ref.~\cite{disintegration} approximated the wall as null). In a particular cosmological model, the area $A$ will be determined by the energy density in the vicinity of our vacuum. If a bubble nucleated in false vacuum is to hit the area $A$, it must form no farther away than the false vacuum horizon length $(G_N \epsilon_A)^{-1/2}$, both in space and time. Therefore, the factor $A/G_N \epsilon_A$ measures the four-volume of false vacuum where bubble nucleations (of either type) lead to collisions. 

In the case of the barnacle, one can make a similar estimate. The relevant decay rate is $\Gamma_b$, the rate per wall three-volume. Thus, we must write down the three-volume of wall that is contained within our past lightcone. In particular, we must recognize that the wall is not null. Indeed, recall that its Lorentzian evolution is  given by a three-dimensional hyperboloid
\begin{equation}
r^2 - t^2 = R_{AC}^2.
\end{equation}
If $A$ denotes the same cross-sectional area as above, it is easy to show that  the desired three-volume is proportional to $R_{AC} A$ (consistently with dimensional analysis). Using $R_{AC} = 3 \sigma_{AC} / \epsilon_{AC}$, the expected number of barnacles in our past lightcone is:
\begin{equation}
N_b \sim \Gamma_b\, \frac{3\sigma_{AC} A}{\epsilon_{AC}}
\label{barnaclecount}
\end{equation}
Comparing eqs.~(\ref{collisioncount}) and (\ref{barnaclecount}), we see that barnacles dominate over bubble collisions if
\begin{equation}
e^{S_{AC}-S_b} \gg \frac{\epsilon_{AC}}{\epsilon_{A}} \frac{K_{AC}}{\sigma_{AC} K_b G_N }
\qquad {\rm and} \qquad
e^{S_{AB}-S_b} \gg \frac{\epsilon_{AC}}{\epsilon_{A}} \frac{K_{AB}}{\sigma_{AC} K_b G_N }
\label{condition} 
\end{equation}
hold. On the left hand sides, the exponents are large and positive, generically of the same order as $S_{AC}$ itself \cite{barnacles}. Thus, violating (\ref{condition}) would be tantamount to stipulating that the magnitudes of the decay rates $\Gamma_{AC}, \Gamma_b$ are controlled as much by the determinant factors $K_{AC}, K_b$ as by the exponentials. This shows that in models where barnacles are of Types III or IV, they are more abundant than bubble collisions.

Barnacles of Types I and II cover half or more of the wall surrounding the region occupied by the true vacuum. Thus, models that give rise to these types of barnacles cannot realize a scenario wherein the walls of our initially spherical bubble create multiple barnacles that produce azimuthal anisotropies in the CMB. Instead, in such models the observable universe is the barnacle itself. The cosmology inside a barnacle is axisymmetric and gives rise to azimuthal features without the need of any collisions. If Planck and other experiments detect the azimuthal signatures discussed in \cite{worldscollide2, flow, polarization, polarization2}, should we infer that we live in a Type I / II barnacle or that we have seen a bubble collision? As an inference problem, the answer depends on the prior probability of each of these scenarios -- i.e. a choice of cosmological measure. 

In any case, it would appear that barnacles are a more robust mechanism for producing axisymmetric cosmological signatures than are bubble collisions. If the underlying parameters place us in a Type III / IV scenario, barnacle nucleations are more frequent than collisions. In a Type I / II scenario, barnacles guarantee the presence of azimuthal effects while bubble collisions require the occurrence rate (\ref{collisioncount}) to be sufficiently high.

A number of possible extensions could place barnacle-related cosmologies on a firmer footing. The barnacle saddle points involve junctions -- loci where three types of walls coalesce. Ref.~\cite{barnacles} does not contain a proof that the requisite solutions to microscopic field equations exist. Further, the analysis of \cite{barnacles} does not include gravity, which significantly affects nucleation rates \cite{cdl}. In Lorentzian signature barnacles grow on the surface of the host wall. It would be interesting to understand what happens when two barnacles collide.

On the other hand, the finding that barnacles are more generic than, but observationally indistinguishable from, bubble collisions is intriguing. I discussed above the relevance of barnacles to the program of using cosmological observations to extract evidence for the string landscape. Beyond that, barnacles should be important for a conceptual understanding of eternal inflation. Because barnacles contain regions where three vacua and three walls coalesce, they lead to nontrivial correlations between which vacua occupy spacelike-separated points. This should have a bearing for qualitative and quantitative studies of eternal inflation such as \cite{frwcft, sss, grainy, symmetree}.

On a more formal level, to my knowledge this paper presents the first exposition of the role of Euclidean saddle points with two or more negative modes. Prior studies \cite{colemanbook, negeig} have specifically emphasized that only saddle points with a single negative mode are relevant to instabilities. This statement holds for decay channels of \emph{empty vacua}, but it does not apply to instabilities of other objects, such as walls created in prior nucleations. The main conclusion of the present paper is that at least in the context of vacuum decay, all saddle points correspond to decay channels. For saddle points whose components cannot be taken arbitrarily far apart, the number of negative directions corresponds to the generation, in which the decay occurs. Thus, a saddle point with $n$ negative modes represents a decay channel of a metastable configuration, which is itself a product of $n-1$ prior decay events.

\section*{Acknowledgements}
I am most grateful to Vijay Balasubramanian for collaboration on the prequel to this paper \cite{barnacles}, for comments on the manuscript and for suggesting the term ``barnacle.'' I also thank Klaus Larjo and Thomas Levi for collaboration on \cite{barnacles} and Ian Affleck, Per Kraus, Ian Moss, and Moshe Rozali for discussions. This paper is dedicated to my wife, Stella Christie. The work was supported by an NSERC discovery grant.

\end{document}